\documentclass[titlepage,12pt]{article}
\usepackage{amsfonts}
\usepackage{geometry}
\usepackage[doublespacing]{setspace}
\usepackage{eucal}
\usepackage{times}

\input{tcilatex}
\begin{document}

\title{Dynamic Retrospective Regression for Functional Data}
\author{Daniel Gervini \\
Department of Mathematical Sciences\\
University of Wisconsin--Milwaukee}
\maketitle

\begin{abstract}
Samples of curves, or functional data, usually present phase variability in
addition to amplitude variability. Existing functional regression methods do
not handle phase variability in an efficient way. In this paper we propose a
functional regression method that incorporates phase synchronization as an
intrinsic part of the model, and then attains better predictive power than
ordinary linear regression in a simple and parsimonious way. The
finite-sample properties of the estimators are studied by simulation. As an
example of application, we analyze neuromotor data arising from a study of
human lip movement.

\end{abstract}

\section{\label{sec:Introduction}Introduction}

Many statistical applications today involve modeling curves as functions of
other curves. For example, the trajectories of CD4 cell counts over time in
HIV patients can be modeled as functions of viral load trajectories (Liang
et al.~2003, Wu and Liang 2004, Wu and M\"{u}ller 2010); gene expression
profiles of insects at the pupal stage can be modeled as functions of gene
expression profiles at the embryonic stage (M\"{u}ller et al.~2008);
trajectories of systolic blood pressure over the years can be predicted to
some extent from trajectories of body mass index (Yao et al.~2005). All of
these examples fall into the relatively new area of functional regression,
or regression methods for functional data.

Functional linear regression, in particular, is a more or less
straightforward extension of multivariate linear regression to the
functional-data framework (Ramsay and Silverman 2005, ch.~16). Recent
developments in functional linear regression have focused on theoretical
aspects such as rates of convergence (Cai and Hall 2006, Hall and Horowitz
2007, Crambes et al.~2009), sparse longitudinal data (Yao et al.~2005), and
interpretability of the estimators (James et al.~2009). But a problem
inherent to functional data that has received little attention in the
regression context is the problem of phase variability.

As a motivating example, consider the data in Malfait and Ramsay (2003). The
authors want to predict lip acceleration using electromyography (EMG) curves
that measure neural activity in the primary muscle that depresses the lower
lip, the depressor labii inferior. A person was asked to repeat the phrase
\textquotedblleft say Bob again\textquotedblright\ a few times, and the lip
movement and associated EMG curve corresponding to the word
\textquotedblleft Bob\textquotedblright\ were recorded. Lip acceleration
curves were obtained by differentiating the smoothed lip trajectories. The
sample curves, time-standardized to 700 msec, are shown in Figure \ref%
{fig:Sample}(a,b). Both samples follow regular patterns, but they show
considerable variability in amplitude and timing of the main features. In
fact, phase variability overwhelms amplitude variability in Figure \ref%
{fig:Sample}(a), to the point that it is hard to tell how many systematic
peaks a typical EMG curve has in the range .3--.7. A pair-by-pair analysis
of the curves shows that the EMG spikes are aligned with certain features of
the acceleration curves; therefore, the \emph{timing} of the EMG spikes (not
just their amplitude) is likely to provide valuable information for
predicting lip acceleration.

\FRAME{ftbpFU}{5.9681in}{3.4791in}{0pt}{\Qcb{Lip Movement Example. (a) EMG
curves; (b) lip acceleration curves; (c) synchronized EMG curves; (d)
synchronized lip acceleration curves.}}{\Qlb{fig:Sample}}{%
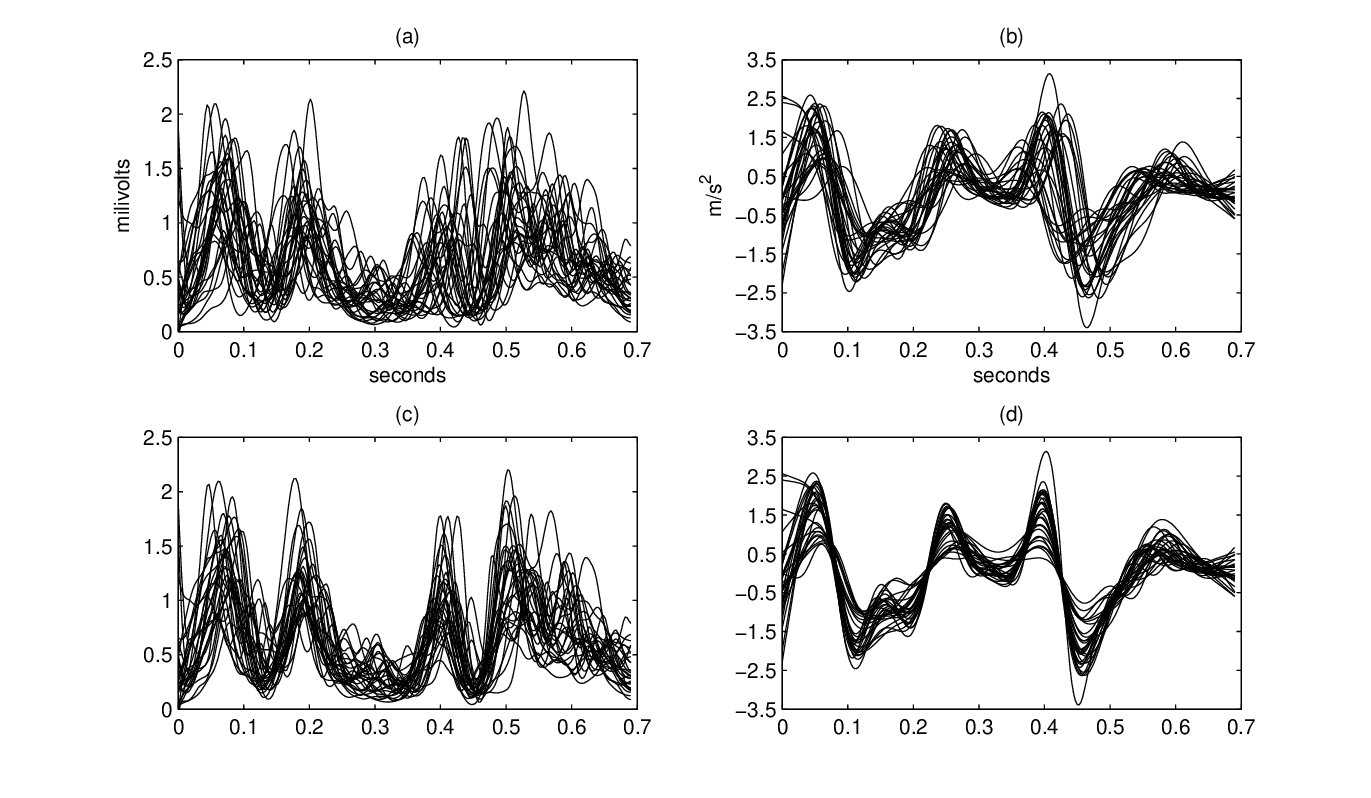}{\special{language "Scientific Word";type
"GRAPHIC";maintain-aspect-ratio TRUE;display "USEDEF";valid_file "F";width
5.9681in;height 3.4791in;depth 0pt;original-width 9.135in;original-height
5.3082in;cropleft "0";croptop "1";cropright "1";cropbottom "0";filename
'samp_and_reg_curves.eps';file-properties "XNPEU";}}

Ordinary functional linear regression does not model phase variability
explicitly. This creates some problems, because phase variability tends to
spread the features of predictor and response curves over wide time ranges
and as a result the regression function becomes very irregular and hard to
interpret. On the other hand, if the curve features are synchronized, a
simpler regression function will provide a good fit to the data.

Several methods of curve synchronization have been proposed over the years.
We can mention Gervini and Gasser (2004, 2005), James (2007), Kneip et
al.~(2000), Kneip and Ramsay (2008), Liu and M\"{u}ller (2004), Ramsay and
Li (1998), Tang and M\"{u}ller (2008, 2009), and Wang and Gasser (1999),
among others. But in a regression context, if covariate and response curves
are synchronized independently it becomes impossible to predict new
(un-warped) response curves from given (un-warped) covariate curves, since
the associated warping functions cannot be predicted.

To address this problem, in this paper we propose a regression method that
incorporates time warping as an intrinsic part of the model. Since we are
going to apply this method to the lip movement data, we will focus on the
retrospective regression model, or \textquotedblleft
historical\textquotedblright\ regression model\ as Malfait and Ramsay (2003)
call it, where $x$ and $y$ are functions of time and for each $t$ we model
the response value $y(t)$ as a function of \textquotedblleft
past\textquotedblright\ covariate values $x(s)$ with $s\leq t$. In addition,
we assume that the sample curves are smooth, as in Figure \ref{fig:Sample}.
Extending this model to sparse and irregular curves is something that will
be addressed in other papers.

\section{\label{sec:Main}Dynamic retrospective regression}

\subsection{Model and estimation}

Let $(x_{1},y_{1}),\ldots ,(x_{n},y_{n})$ be a sample of functions, where $%
x_{i}$ is the covariate curve and $y_{i}$ the response curve. We assume $%
x_{i}$ and $y_{i}$ are square-integrable functions on a common\ interval $%
[a,b]$. A linear predictor of $y_{i}(t)$ based on $x_{i}$ has the form 
\begin{equation}
L(t;x_{i},\alpha ,\beta )=\alpha (t)+\int_{a}^{b}\beta (s,t)x_{i}(s)ds,
\label{eq:L}
\end{equation}%
where $\alpha $ is the intercept function and $\beta $ the slope function.
However, (\ref{eq:L}) employs the whole trajectory $\{x_{i}(s):s\in \lbrack
a,b]\}$ to predict $y_{i}(t)$, including \textquotedblleft
future\textquotedblright\ observations $x_{i}(s)$ with $s>t$. In many
applications this is not reasonable. For example, for the lip movement data
in Figure \ref{fig:Sample} it is clear that future neural activity cannot
have an influence on past lip movement; therefore, prediction of $y_{i}(t)$
must be based only on the partially observed curves $\{x_{i}(s):s\in \lbrack
a,t]\}$. Then instead of (\ref{eq:L}) we will use 
\begin{equation}
RL(t;x_{i},\alpha ,\beta )=\alpha (t)+\int_{a}^{t}\beta (s,t)x_{i}(s)ds,
\label{eq:RL}
\end{equation}%
which can be seen as a particular case of (\ref{eq:L}) under the constraint $%
\beta (s,t)=0$ for $s>t$. This model is called \textquotedblleft historical
linear model\textquotedblright\ by Malfait and Ramsay (2003), although we
prefer the denomination \textquotedblleft retrospective linear
model\textquotedblright .

As explained in the Introduction, ordinary functional linear regression
works best for synchronized curves. Suppose, then, that for each pair $%
(x_{i},y_{i})$ we have a warping function $w_{i}:[a,b]\rightarrow \lbrack
a,b]$, that is, a strictly monotone increasing function that satisfies $%
w_{i}(a)=a$ and $w_{i}(b)=b$. Let $\tilde{x}_{i}=x_{i}\circ w_{i}$ and $%
\tilde{y}_{i}=y_{i}\circ w_{i}$ be the warped curves; then we apply (\ref%
{eq:RL}) to $(\tilde{x}_{i},\tilde{y}_{i})$ rather than $(x_{i},y_{i})$, and
define the \emph{dynamic functional predictor} of $y_{i}(t)$ as $%
RL(w_{i}^{-1}(t);\tilde{x}_{i},\alpha ,\beta )$, obtaining 
\begin{equation}
\hat{y}_{i}(t)=\alpha (w_{i}^{-1}(t))+\int_{a}^{w_{i}^{-1}(t)}\beta
(s,w_{i}^{-1}(t))x_{i}(w_{i}(s))ds.  \label{eq:DynL}
\end{equation}%
Note that the same warping function $w_{i}$ is used for $x_{i}$ and $y_{i}$;
this is reasonable for the type of applications we have in mind. Using a
common warping function preserves the retrospective property of the model:
the integral in (\ref{eq:DynL}) only involves values of $x_{i}(w_{i}(s))$
with $s\leq w_{i}^{-1}(t)$, or equivalently, $x_{i}(s)$ with $s\leq t$.

The estimators of $\alpha $, $\beta $, and the $w_{i}$s can be obtained by
functional least squares, minimizing 
\begin{equation}
\sum_{i=1}^{n}\left\Vert y_{i}\circ w_{i}-RL(\cdot ;x_{i}\circ w_{i},\alpha
,\beta )\right\Vert ^{2}  \label{eq:LS}
\end{equation}%
with respect to $\alpha $, $\beta $ and the $w_{i}$s, where $\left\Vert
f\right\Vert =\{\int_{a}^{b}f^{2}(t)dt\}^{1/2}$ is the usual $L^{2}([a,b])$%
-norm. Note that for given $\beta $ and $w_{i}$s, the $\alpha $ that
minimizes (\ref{eq:LS}) is 
\begin{equation}
\hat{\alpha}(t)=\overline{\tilde{y}}(t)-\int_{a}^{t}\beta (s,t)\overline{%
\tilde{x}}(s)ds,  \label{eq:alpha_hat}
\end{equation}%
so we can re-write (\ref{eq:LS}) as 
\begin{equation}
\sum_{i=1}^{n}\int_{a}^{b}\left[ \tilde{y}_{i}(t)-\overline{\tilde{y}}%
(t)-\int_{a}^{t}\beta (s,t)\{\tilde{x}_{i}(s)-\overline{\tilde{x}}(s)\}ds%
\right] ^{2}dt,  \label{eq:LS-centered}
\end{equation}%
eliminating the intercept $\alpha $.

The estimation of $\beta $ has to be done with care in order to avoid
identifiability issues. To understand this problem, consider again the
general linear predictor (\ref{eq:L}). For any function $\gamma $ such that $%
\int_{a}^{b}\gamma (s,t)x_{i}(s)ds=0$ for all $t$ and all $i$, it is clear
that $L(t;x_{i},\alpha ,\beta )=L(t;x_{i},\alpha ,\beta +\gamma )$ for all $%
t $ and all $i$, so (\ref{eq:L}) cannot distinguish between $\beta $ and $%
\beta +\gamma $. Since the space spanned by the $x_{i}$s has dimension at
most $n$, there is always going to be an infinite number of $\gamma $s for
which this occurs. The usual way to deal with this identifiability issue is
to reduce the space of possible $\beta $s, so that the only $\gamma $ that
satisfies $\int_{a}^{b}\gamma (s,t)x_{i}(s)ds=0$ for all $t$ and all $i$ in
the reduced space is $\gamma \equiv 0$. An efficient way to do this is to
use the tensor-product space of the principal components of the $x_{i}$s and
the $y_{i}$s (e.g.~as in M\"{u}ller \emph{et al.}, 2008), which is the
functional equivalent of principal-component regression.

We briefly remind the reader what the functional principal components are. A
continuous covariance function $\rho (s,t)=\mathrm{cov}\{x(s),x(t)\}$ admits
the decomposition $\rho (s,t)=\sum_{k}\lambda _{k}\phi _{k}(s)\phi _{k}(t)$,
where the $\phi _{k}$s are orthogonal functions in $L^{2}([a,b])$ and the $%
\lambda _{k}$s are non-increasing positive scalars (the sequence may be
finite or infinite, but in either case $\sum_{k}\lambda _{k}<\infty $). This
is known as Mercer's Theorem (Gohberg \emph{et al.}, 2003). The $\phi _{k}$s
are eigenfunctions of $\rho $, i.e.~they satisfy $\int_{a}^{b}\rho (s,t)\phi
_{k}(s)ds=\lambda _{k}\phi _{k}(t)$ for all $t$. The $\phi _{k}$s are called
the principal components of the $x$-space, since they are the functional
equivalents of the multivariate principal components. In a similar way one
obtains the principal components of the $y$-space, say $\{\psi _{l}\}$. To
estimate the regression function $\beta $, one chooses the leading principal
components of each space, say $\{\phi _{1},\ldots ,\phi _{p}\}$ and $\{\psi
_{1},\ldots ,\psi _{q}\}$, and sets 
\begin{equation}
\beta (s,t)=\sum_{k=1}^{p}\sum_{l=1}^{q}b_{kl}\phi _{k}(s)\psi _{l}(t).
\label{eq:exp_beta}
\end{equation}%
The coefficients $b_{kl}$ of $\beta $ are estimated by least squares. This
procedure can be adapted for the retrospective linear predictor (\ref{eq:RL}%
) in a straightforward way, since the estimation of the principal components
does not change.

Going back to the problem of minimizing (\ref{eq:LS-centered}), we proceed
as follows. First note that the $\phi _{k}$s and the $\psi _{l}$s are now
the principal components of the warped functions $\{\tilde{x}_{i}\}$ and $\{%
\tilde{y}_{i}\}$, respectively, so we cannot estimate them separately in a
preliminary step because they depend on the warping functions $\{w_{i}\}$,
which are themselves estimated in the process. So we minimize (\ref%
{eq:LS-centered}) with respect to $\mathbf{B}=[b_{kl}]$ and the $w_{i}$s,
subject to the conditions (\ref{eq:exp_beta}) and 
\begin{eqnarray}
\int_{a}^{b}\rho _{\tilde{x}}(s,t)\phi _{k}(s)ds &=&\lambda _{k}\phi
_{k}(t),\ \ k=1,\ldots ,p,  \label{eq:Phi} \\
\int_{a}^{b}\rho _{\tilde{y}}(s,t)\psi _{k}(s)ds &=&\xi _{k}\psi _{k}(t),\ \
k=1,\ldots ,q,  \label{eq:Psi}
\end{eqnarray}%
where 
\begin{eqnarray*}
\rho _{\tilde{x}}(s,t) &=&\frac{1}{n}\sum_{i=1}^{n}\{\tilde{x}_{i}(s)-%
\overline{\tilde{x}}(s)\}\{\tilde{x}_{i}(t)-\overline{\tilde{x}}(t)\}, \\
\rho _{\tilde{y}}(s,t) &=&\frac{1}{n}\sum_{i=1}^{n}\{\tilde{y}_{i}(s)-%
\overline{\tilde{y}}(s)\}\{\tilde{y}_{i}(t)-\overline{\tilde{y}}(t)\}.
\end{eqnarray*}%
In addition, we assume that the $\phi _{k}$s and the $\psi _{l}$s are
orthonormal and that the sequences $\{\lambda _{k}\}$ and $\{\xi _{k}\}$ are
positive and non-increasing. For identifiability of the warping functions,
we also add the constraint $\bar{w}(t)\equiv t$.

It is convenient to model the functional parameters $\{\phi _{k}\}$, $\{\psi
_{l}\}$ and $\{w_{i}\}$ using splines or similar basis functions, because
this reduces the functional minimization problem to a more familiar
multivariate minimization problem. Let $\mathbf{\gamma }(t)=(\gamma
_{1}(t),\ldots ,\gamma _{\nu }(t))$ be a spline basis (or some other system)
in $L^{2}([a,b])$; then we assume $\phi _{k}(t)=\sum_{j=1}^{\nu
}c_{kj}\gamma _{j}(t)$ and $\psi _{l}(t)=\sum_{j=1}^{\nu }d_{lj}\gamma
_{j}(t)$ for coefficient vectors $\mathbf{c}_{k}$ and $\mathbf{d}_{l}$. The
regression slope can then be expressed as 
\[
\beta (s,t)=\mathbf{\gamma }(s)^{T}\mathbf{C}^{T}\mathbf{BD\gamma }%
(t)I\{s\leq t\} 
\]%
and the functional constraints (\ref{eq:Phi}) and (\ref{eq:Psi}) turn into
parametric constraints 
\begin{eqnarray}
\mathbf{\Omega }_{\tilde{x}}\mathbf{C} &=&\mathbf{J}_{0}\mathbf{C\Lambda },
\label{eq:constr_C} \\
\mathbf{\Omega }_{\tilde{y}}\mathbf{D} &=&\mathbf{J}_{0}\mathbf{D\Xi },
\label{eq:constr_D}
\end{eqnarray}%
where $\mathbf{\Omega }_{\tilde{x}}=\iint \rho _{\tilde{x}}(s,t)\mathbf{%
\gamma }(s)\mathbf{\gamma }(t)^{T}\ ds\ dt$, $\mathbf{\Omega }_{\tilde{y}%
}=\iint \rho _{\tilde{y}}(s,t)\mathbf{\gamma }(s)\mathbf{\gamma }(t)^{T}\
ds\ dt$, $\mathbf{J}_{0}=\int \mathbf{\gamma }(t)\mathbf{\gamma }(t)^{T}dt$, 
$\mathbf{C}=[\mathbf{c}_{1},\ldots ,\mathbf{c}_{p}]$, $\mathbf{D}=[\mathbf{d}%
_{1},\ldots ,\mathbf{d}_{q}]$, $\mathbf{\Lambda }=\mathrm{diag}(\lambda
_{1},\ldots ,\lambda _{p})$ and $\mathbf{\Xi }=\mathrm{diag}(\xi _{1},\ldots
,\xi _{q})$. Note that $\mathbf{\Omega }_{\tilde{x}}$ and $\mathbf{\Omega }_{%
\tilde{y}}$ are functions of the $w_{i}$s via $\rho _{\tilde{x}}$ and $\rho
_{\tilde{y}}$, but we omit this in the notation for simplicity. In addition,
we also have the orthogonality conditions $\mathbf{C}^{T}\mathbf{J}_{0}%
\mathbf{C}=\mathbf{I}_{p}$ and $\mathbf{D}^{T}\mathbf{J}_{0}\mathbf{D}=%
\mathbf{I}_{q}$.

Parameterizing the warping functions is more complicated due to their
monotonicity. One possibility is to model the $w_{i}$s as B-spline functions
with monotone increasing coefficients, which guarantees that the $w_{i}$s
are monotone increasing (Brumback and Lindstrom, 2004); the boundary
conditions $w_{i}(a)=a$ and $w_{i}(b)=b$ and the identifiability condition $%
\bar{w}(t)\equiv t$ can be expressed as linear constraints on the
coefficients. Another possibility is to use the family of smooth\ monotone
transformations (Ramsay and Li, 1998), where $\log \{w_{i}^{\prime }(t)\}$
is modeled as an unconstrained B-spline function and $w_{i}(t)$ is computed
by integration; if $\mathbf{\theta }_{i}\in \mathbb{R}^{r}$ are the spline
coefficients of $\log \{w_{i}^{\prime }(t)\}$, a convenient identifiability
condition is the restriction $\mathbf{\bar{\theta}}=\mathbf{0}$, which
approximately implies $\bar{w}(t)\equiv t$. A third possibility, which is
the one we prefer in this paper, is to model the $w_{i}$s as monotone
interpolating cubic Hermite splines (Fritsch and Carlson, 1980). This family
is specified by a vector of knots $\mathbf{\tau }_{0}\in \mathbb{R}^{r}$ in $%
(a,b)$ and each $w_{i}$ is determined by a corresponding vector $\mathbf{%
\tau }_{i}$ such that $w_{i}(\mathbf{\tau }_{0})=\mathbf{\tau }_{i}$. The $%
\mathbf{\tau }_{i}$s then become the parameters that determine $w_{i}$. The
strategy, when using Hermite splines, is to place the knots $\mathbf{\tau }%
_{0}$ at locations of interest, such as the (approximate) average location
of peaks and valleys. For example, for the lip movement data in Figure \ref%
{fig:Sample} a reasonable choice would be $\mathbf{\tau }_{0}=(.1,.2,.4,.5)$%
, corresponding to the approximate average location of the peaks of the $%
x_{i}$s. This way we obtain warping flexibility at the features of interest
with a low-dimensional family of warping functions, since the $\mathbf{\tau }%
_{i}$s can take any value as long as $a<\tau _{i1}<\cdots <\tau _{ir}<b$.
One technicality: due to this monotonicity restriction, it is
computationally more convenient to use the Jupp transforms (Jupp, 1978) of
the $\mathbf{\tau }_{i}$s, 
\[
\theta _{ij}=\log \{(\tau _{i,j+1}-\tau _{ij})/(\tau _{ij}-\tau
_{i,j-1})\},\ \ j=1,\ldots ,r, 
\]%
as parameters, because the $\mathbf{\theta }_{i}$s are unconstrained
vectors. The identifiability condition $\mathbf{\bar{\theta}}=\mathbf{\theta 
}_{0}$ approximately implies that $\mathbf{\bar{\tau}}=\mathbf{\tau }_{0}$
and therefore $\bar{w}(t)\equiv t$. More details about monotone Hermite
splines are given in the Technical Supplement.

The minimization of (\ref{eq:LS-centered}) has thus become a multivariate
constrained minimization problem on the parameters $\mathbf{B}$, $\mathbf{C}$%
, $\mathbf{D}$, $\mathbf{\Lambda }$, $\mathbf{\Xi }$, and $\mathbf{\theta }%
_{1}$,\ldots , $\mathbf{\theta }_{n}$, which can be solved via standard
optimization methods (see e.g.~Nocedal and Wright, 2006, ch.~15). In our
Matlab programs we use the interior-point algorithm as implemented in
Matlab's function \textquotedblleft fmincon\textquotedblright . This type of
algorithms converge only to a local minimum, so it is important to select a
good starting point to increase the chances of actually finding the global
minimum, or at least a \textquotedblleft good\textquotedblright\ local
solution. One approach we have found successful is to do a quick (separate)
synchronization of the $x_{i}$s and the $y_{i}$s and use the resulting
principal components as initial estimators of the $\phi _{k}$s and the $\psi
_{l}$s, and the warping parameters of either sample as initial $\mathbf{%
\theta }_{i}$s. Another alternative is to try several random starting
points, but this is much more time consuming.

Once the estimators $\hat{\alpha}$, $\hat{\beta}$ and $\hat{w}_{1},\ldots ,%
\hat{w}_{n}$ have been obtained, it is possible to use to predict a response
function $y_{n+1}$ for a given covariate function $x_{n+1}$. The natural
predictor of $y_{n+1}$ given $x_{n+1}$ is $\hat{y}_{n+1}(t)=RL(\hat{w}%
_{n+1}^{-1}(t);\tilde{x}_{n+1},\hat{\alpha},\hat{\beta})$, but $\hat{w}%
_{n+1} $ cannot be obtained by minimizing the integrated squared residual
because that involves the unobserved response $y_{n+1}$. Instead, $\hat{w}%
_{n+1}$ can be obtained from $x_{n+1}$ alone by synchronizing $x_{n+1}$ to
the mean of the warped $x_{i}$s: 
\begin{equation}
\hat{w}_{n+1}=\arg \min_{w}\left\Vert x_{n+1}\circ w-\overline{\tilde{x}}%
\right\Vert ^{2}.  \label{eq:wLS}
\end{equation}%
Note that $\overline{\tilde{x}}$ here is fixed, so the \textquotedblleft
pinching\textquotedblright\ or \textquotedblleft
overwarping\textquotedblright\ problem associated with least-squares
registration [discussed in Ramsay and Silverman (2005, ch.~7.6) and Kneip
and Ramsay (2008)] will not be a serious issue.

\subsection{Selection of meta-parameters}

In addition to the parameters estimated by least squares, there are some
meta-parameters that also need to be specified. For example, the number and
placement of knots of the spline bases used for the $\phi _{k}$s, the $\psi
_{l}$s and the $w_{i}$s, and most importantly $p$ and $q$, the number of
principal components to be included in (\ref{eq:exp_beta}). The simplest
approach would be to minimize computationally simple criteria such as the
\textquotedblleft generalized cross validation\textquotedblright\ criterion 
\[
\mathrm{GCV}(p,q)=\mathrm{MSE}(p,q)/(1-pq/n)^{2} 
\]%
(Wahba, 1990) or the \textquotedblleft corrected Akaike
criterion\textquotedblright\ 
\[
\mathrm{AICC}(p,q)=\mathrm{MSE}(p,q)\exp \{1+2(pq+1)/(n-pq-2)\} 
\]%
(Hurvitz \emph{et al.}, 1998), where $\mathrm{MSE}=n^{-1}\sum_{i=1}^{n}\Vert 
\tilde{y}_{i}^{\ast }-\widehat{\tilde{y}_{i}^{\ast }}\Vert ^{2}$ and $\tilde{%
y}_{i}^{\ast }(t)=\tilde{y}_{i}(t)-\overline{\tilde{y}}(t)$. Unfortunately
these criteria did not perform well in our simulations, so a more
computationally complex approach such as $k$-fold cross-validation (Hastie 
\emph{et al.}, 2009) needs to be explored.

Regarding the spline bases for the $\phi _{k}$s and the $\psi _{l}$s, we
note that since the method \textquotedblleft borrows
strength\textquotedblright\ across curves to estimate the coefficients $\{%
\mathbf{c}_{k}\}$ and $\{\mathbf{d}_{k}\}$, the spline dimension $\nu $ can
be relatively large and the resulting $\phi _{k}$s and the $\psi _{l}$s will
still be reasonably regular. Thus we simply take a fairly large number of
equally spaced points in $(a,b)$ as knots. If necessary, roughness penalty
terms can be added to control the regularity of the $\phi _{k}$s and the $%
\psi _{l}$s. In our implementation, we do that by adding the penalty to the
constraints (as in Silverman, 1996), substituting $\mathbf{J}_{0}$ in (\ref%
{eq:constr_C}), (\ref{eq:constr_D}) and in the orthogonality constraints by $%
\mathbf{J}_{0}+\eta \mathbf{J}_{2}$, where $\mathbf{J}_{2}=\int \mathbf{%
\gamma }^{\prime \prime }(t)\mathbf{\gamma }^{\prime \prime }(t)^{T}dt$ and $%
\eta $ is a roughness-penalty parameter, that in practice is chosen
subjectively.

Regarding the warping functions, the approach to follow will depend on the
warping family. If interpolating Hermite splines are used, a small number of
knots $r$ placed nearby the salient landmarks usually provide ample warping
flexibility, and $r$ can be chosen by cross-validation along with $p$ and $q$
within a small range of triplets $(p,q,r)$. If smooth monotone
transformations are used, for which spline knots are not identified with
meaningful landmarks, a better approach is to use several equally-spaced
knots and add the roughness-penalty term $\eta \sum_{i=1}^{n}\int \{(\log
w_{i}^{\prime })^{\prime }\}^{2}$ to the objective function, as in Ramsay
and Li (1998); in that case the smoothing parameter $\eta $ must be chosen
with care, because it determines the effective dimension of the warping
space and therefore there is going to be an interplay between $p$, $q$ and $%
\eta $.

\section{\label{sec:Simulations}Simulations}

We ran two sets of simulations to assess the performance of the proposed
method. The first set was designed to compare the dynamic regression
estimator with the ordinary functional least squares estimator, to determine
to what extent the estimators are able to reconstruct the true regression
function $\beta $. The data was generated as follows. We generated $\tilde{x}%
_{i}$s following the shape-invariant model $\tilde{x}%
_{i}(s)=z_{i}e^{-30(s-.4)^{2}}$ with $z_{i}$ i.i.d.~$N(1,.2^{2})$, which is
a one-component model with $\mu _{\tilde{x}}(s)=e^{-30(s-.4)^{2}}$ and $\phi
_{1}=\mu _{\tilde{x}}/\Vert \mu _{\tilde{x}}\Vert $. The $\tilde{y}_{i}$s
were generated as 
\begin{equation}
\tilde{y}_{i}(t)=\int_{0}^{t}\beta (s,t)\tilde{x}_{i}(s)ds+\varepsilon
_{i}(t)  \label{eq:y_tilde_sim}
\end{equation}%
with $\beta (s,t)=5e^{-50\{(s-.4)^{2}+(t-.6)^{2}\}}$, and $\varepsilon
_{i}(t)=u_{i}\sin (6\pi t)$ with $u_{i}$ i.i.d.~$N(0,\sigma ^{2})$. We
considered two possibilities: a model without random error, where $\sigma =0$%
, and a model with $\sigma =.10$. The effect of the regression function $%
\beta $ is, basically, to shift the peak from $.4$ to $.6$. Note that (\ref%
{eq:y_tilde_sim}) induces a one-component model for the $\tilde{y}_{i}$s in
the $\sigma =0$ case, with $\mu _{\tilde{y}}(t)=\int_{0}^{t}\beta (s,t)\mu _{%
\tilde{x}}(s)ds$ and $\psi _{1}=\mu _{\tilde{y}}/\Vert \mu _{\tilde{y}}\Vert 
$; whereas it induces a two-component model in the $\sigma =.10$ case.

Regarding the warping functions, we also considered two situations: data
without warping, where $(x_{i},y_{i})=(\tilde{x}_{i},\tilde{y}_{i})$, and
warped data $(x_{i},y_{i})=(\tilde{x}_{i}\circ w_{i}^{-1},\tilde{y}_{i}\circ
w_{i}^{-1})$, with warping functions $w_{i}(t)=(e^{a_{i}t}-1)/(e^{a_{i}}-1)$
where the $a_{i}$s are uniformly distributed in $[-1,1]$. Ten random pairs $%
(x_{i},y_{i})$ of the latter case are shown in Figure \ref%
{fig:Simulated_data}(a,b) for illustration. Two sample sizes were
considered: $n=50$ and $n=100$. We will refer to this model as
\textquotedblleft Model 1\textquotedblright .

\FRAME{ftbpFU}{6.1514in}{4.5558in}{0pt}{\Qcb{Simulated data. Ten
illustrative sample curves $(x_{i},y_{i})$ for Model 1 [(a) covariates, (b)
responses] and Model 2 [(c) covariates, (d) responses].}}{\Qlb{%
fig:Simulated_data}}{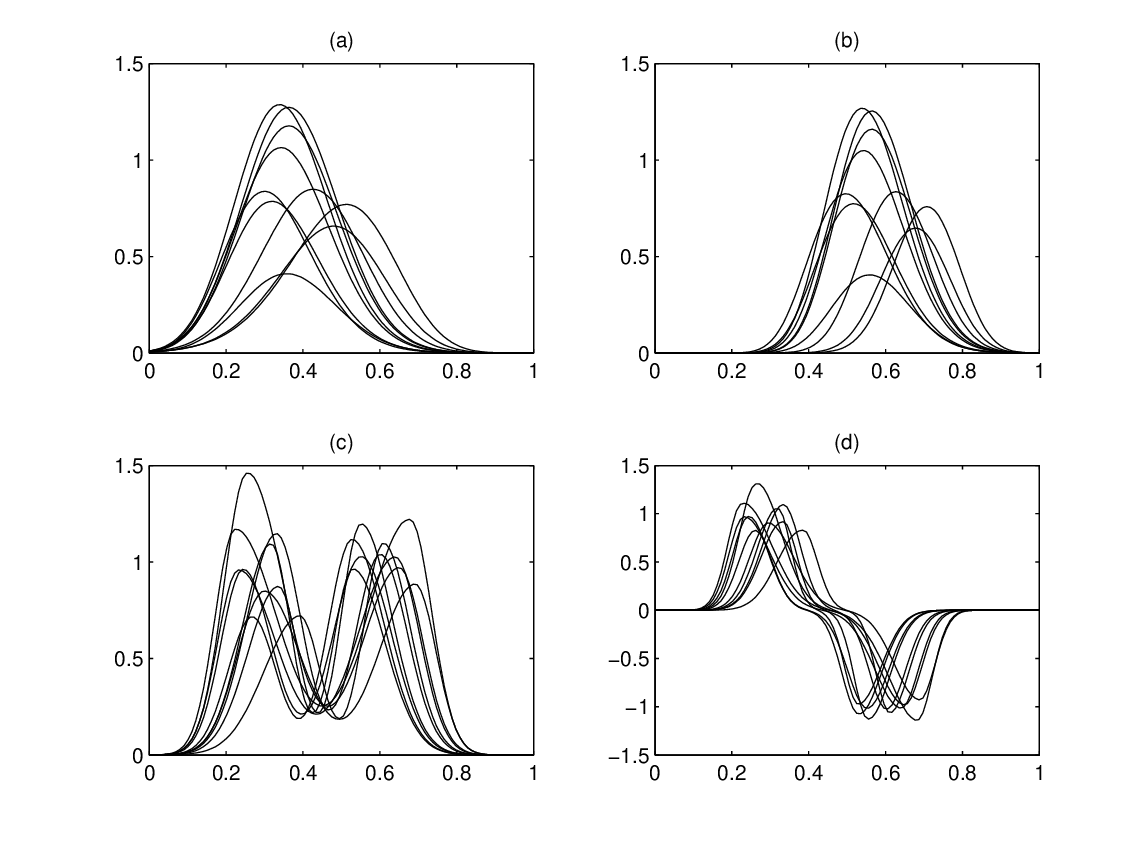}{\special{language "Scientific
Word";type "GRAPHIC";maintain-aspect-ratio TRUE;display "USEDEF";valid_file
"F";width 6.1514in;height 4.5558in;depth 0pt;original-width
9.6945in;original-height 3.0251in;cropleft "0";croptop "1";cropright
"1";cropbottom "0";filename 'Simulated_data.eps';file-properties "XNPEU";}}

For this set of simulations we implemented the dynamic functional regression
estimator with Hermite splines, using a single knot at $\tau _{0}=.5$. The $%
\phi _{k}$s and $\psi _{l}$s were modeled as cubic B-splines with equally
spaced knots; two cases were considered: four and nine knots, giving $\nu =8$
and $\nu =13$ respectively. The same spline bases were used for the $\phi
_{k}$s and $\psi _{l}$s of the ordinary least squares estimator. Regarding
the choice of dimensions $(p,q)$, we considered four combinations: $(1,1)$, $%
(1,2)$, $(2,1)$, and $(2,2)$. Given that the true $\tilde{x}_{i}$s are
one-dimensional and the true $\tilde{y}_{i}$s are either one-dimensional
(when $\sigma =0$) or two-dimensional (when $\sigma =.10$), we expect the
optimal estimators to correspond to models $(1,1)$ and $(1,2)$, respectively.

\begin{table}[tbp] \centering%
\begin{tabular}{ccccccccc}
& \multicolumn{8}{c}{Model without warping} \\ 
& \multicolumn{4}{c}{4 knots} & \multicolumn{4}{c}{9 knots} \\ 
& \multicolumn{2}{c}{$n=50$} & \multicolumn{2}{c}{$n=100$} & 
\multicolumn{2}{c}{$n=50$} & \multicolumn{2}{c}{$n=100$} \\ 
$(p,q)$ & D & L & D & L & D & L & D & L \\ 
$(1,1)$ & .074 & .066 & .072 & .066 & .142 & .062 & .115 & .062 \\ 
$(1,2)$ & .078 & .066 & .079 & .066 & .129 & .063 & .120 & .063 \\ 
$(2,1)$ & .070 & .103 & .070 & .101 & .130 & .219 & .149 & .222 \\ 
$(2,2)$ & .069 & .138 & .092 & .131 & .123 & .240 & .128 & .231 \\ 
AICC & .071 & .131 & .069 & .129 & .121 & .238 & .103 & .236 \\ 
&  &  &  &  &  &  &  &  \\ 
& \multicolumn{8}{c}{Model with warping} \\ 
& \multicolumn{4}{c}{4 knots} & \multicolumn{4}{c}{9 knots} \\ 
& \multicolumn{2}{c}{$n=50$} & \multicolumn{2}{c}{$n=100$} & 
\multicolumn{2}{c}{$n=50$} & \multicolumn{2}{c}{$n=100$} \\ 
$(p,q)$ & D & L & D & L & D & L & D & L \\ 
$(1,1)$ & .085 & .650 & .082 & .656 & .134 & .593 & .132 & .598 \\ 
$(1,2)$ & .082 & .647 & .080 & .653 & .144 & .595 & .143 & .600 \\ 
$(2,1)$ & .148 & .669 & .175 & .672 & .136 & .612 & .177 & .614 \\ 
$(2,2)$ & .158 & .522 & .317 & .522 & .209 & .506 & .287 & .505 \\ 
AICC & .203 & .522 & .349 & .522 & .207 & .506 & .295 & .505%
\end{tabular}%
\caption{Simulation results for Model 1, case $\sigma = 0$ (no error term). Mean integrated absolute errors
of the slope estimators $\hat{\beta}$ are given, for dynamic regression (D)
and ordinary linear regression (L).}\label{tab:Sim_Mod1_no_err}%
\end{table}%

\begin{table}[tbp] \centering%
\begin{tabular}{ccccccccc}
& \multicolumn{8}{c}{Model without warping} \\ 
& \multicolumn{4}{c}{4 knots} & \multicolumn{4}{c}{9 knots} \\ 
& \multicolumn{2}{c}{$n=50$} & \multicolumn{2}{c}{$n=100$} & 
\multicolumn{2}{c}{$n=50$} & \multicolumn{2}{c}{$n=100$} \\ 
$(p,q)$ & D & L & D & L & D & L & D & L \\ 
$(1,1)$ & .112 & .124 & .110 & .115 & .171 & .140 & .165 & .127 \\ 
$(1,2)$ & .094 & .076 & .090 & .072 & .145 & .074 & .152 & .070 \\ 
$(2,1)$ & .105 & .271 & .132 & .236 & .180 & .513 & .198 & .561 \\ 
$(2,2)$ & .373 & .193 & .342 & .172 & .296 & .256 & .310 & .245 \\ 
AICC & .333 & .083 & .337 & .072 & .276 & .076 & .317 & .069 \\ 
&  &  &  &  &  &  &  &  \\ 
& \multicolumn{8}{c}{Model with warping} \\ 
& \multicolumn{4}{c}{4 knots} & \multicolumn{4}{c}{9 knots} \\ 
& \multicolumn{2}{c}{$n=50$} & \multicolumn{2}{c}{$n=100$} & 
\multicolumn{2}{c}{$n=50$} & \multicolumn{2}{c}{$n=100$} \\ 
$(p,q)$ & D & L & D & L & D & L & D & L \\ 
$(1,1)$ & .118 & .647 & .114 & .650 & .161 & .589 & .156 & .596 \\ 
$(1,2)$ & .100 & .643 & .097 & .646 & .143 & .591 & .174 & .597 \\ 
$(2,1)$ & .172 & .668 & .228 & .671 & .165 & .608 & .205 & .610 \\ 
$(2,2)$ & .480 & .529 & .516 & .526 & .399 & .518 & .378 & .514 \\ 
AICC & .343 & .529 & .531 & .526 & .325 & .518 & .375 & .514%
\end{tabular}%
\caption{Simulation results for Model 1, case $\sigma = .10$. Mean integrated absolute errors
of the slope estimators $\hat{\beta}$ are given, for dynamic regression (D)
and ordinary linear regression (L).}\label{tab:Sim_Mod1_with_err}%
\end{table}%

We would also expect the GCV or the AICC criteria to choose these models as
optimal, if they were useful for model selection. Tables \ref%
{tab:Sim_Mod1_no_err} and \ref{tab:Sim_Mod1_with_err} report mean integrated
absolute errors, $\mathrm{MIAE}(\hat{\beta})=\mathrm{E}\{\iint |\hat{\beta}%
(s,t)-\beta (s,t)|~ds~dt\}$, based on 300 Monte Carlo replications, for $%
\sigma =0$ and $\sigma =.10$ respectively. The $\mathrm{MIAE}$s of the
models selected by AICC and GCV were very similar, so we only report the
results for AICC. We see that in the absence of warping the dynamic
regression estimator is comparable to ordinary least squares, so nothing is
lost by using a more complex estimator. But in presence of warping the
dynamic estimator is clearly better. We know that warping distorts the
principal component estimators and, as a consequence, the ordinary least
squares estimator cannot produce a good estimator of $\beta $ unless too
many components are used, and in that case overfitting is a problem. On the
other hand, the dynamic estimator successfully recovers the principal
components $\phi _{k}$ and $\psi _{l}$ of the $\tilde{x}_{i}$s and $\tilde{y}%
_{i}$s, and therefore it provides an accurate estimator of $\beta $,
especially for the optimal models $(p,q)=(1,1)$ and $(p,q)=(1,2)$.
Unfortunately GCV and AICC do not provide very useful guidance for model
selection, judging from their $\mathrm{MIAE}$s, so alternative procedures
like $k$-fold cross-validation should be explored. We did not study the
performance of $k$-fold cross-validation by simulation but it did prove
useful for the Lip Movement data analysis in Section \ref{sec:Example}.

We also ran a second set of simulations where we varied the dimension of the
warping spaces used for estimation. The data was generated from a more
complex model that we will call \textquotedblleft Model 2\textquotedblright
. The warped covariates $\{\tilde{x}_{i}\}$ followed a two-component model $%
\tilde{x}_{i}(s)=\mu _{\tilde{x}}(s)+\sum_{k=1}^{2}z_{ik}\phi _{k}(s)$ with $%
\mu _{\tilde{x}}(s)=e^{-100(s-.3)^{2}}+e^{-100(s-.6)^{2}}$ and each $\phi
_{k}$ proportional to a peak (more specifically, if $%
g_{1}(s)=e^{-100(s-.3)^{2}}$ and $g_{2}(s)=e^{-100(s-.6)^{2}}$, we took $%
\phi _{1}=g_{1}/\Vert g_{1}\Vert $ and $\phi _{2}=c(g_{2}-\langle \phi
_{1},g_{2}\rangle \phi _{1})$ with $c$ a normalizing constant). The $z_{ik}$%
s were i.i.d.~$N(0,.07^{2})$ and $N(0,.05^{2})$, respectively. As regression
slope we took $\beta (s,t)=\{\phi _{1}(s)\psi _{1}(t)+\phi _{2}(s)\psi
_{2}(t)\}\mathbb{I}\{s\leq t\}$, with $\psi _{1}=\phi _{1}$ and $\psi
_{2}=-\phi _{2}$; the mean of the warped responses $\{\tilde{y}_{i}\}$ was
set as $\mu _{\tilde{y}}(t)=e^{-100(t-.3)^{2}}-e^{-100(t-.6)^{2}}$. So the $%
\tilde{y}_{i}$s have a peak and a valley; the height of the peak is
proportional to the height of the first peak of $\tilde{x}_{i}$, and the
depth of the valley is proportional to the height of the second peak of $%
\tilde{x}_{i}$. No random error $\varepsilon _{i}(t)$ was used for Model 2,
since the results for Model 1 were similar for models with or without random
error. The pair $(\tilde{x}_{i},\tilde{y}_{i})$ was then warped with a $%
w_{i}\left( t\right) $ that had two independent warping knots, one at each
peak. Specifically, we generated $\tau _{i1}\sim U(.2,.4)$ and $\tau
_{i2}\sim U(.5,.7)$ independently and constructed a piecewise linear $%
w_{i}(t)$ such that $w_{i}(0)=0$, $w_{i}(.3)=\tau _{i1}$, $w_{i}(.6)=\tau
_{i2}$ and $w_{i}(1)=1$. A sample of ten pairs $(x_{i},y_{i})$ is shown in
Figure \ref{fig:Simulated_data}(c,d) for illustration. We generated samples
of size $n=50$ and 300 replications were run.

We compared the performance of dynamic functional regression estimators with
two different families of warping functions: Hermite splines and smooth
monotone functions (Ramsay and Li, 1998). We considered three knot sequences 
$\mathbf{\tau }_{0}$ of increasing dimensions: $.50$, $(.33,.66)$, and $%
(.25,.50,.75)$. For Hermite splines we did not penalized the roughness of
the $w_{i}$s, but for smooth monotone functions we did, since the algorithm
tended to produce degenerate warping functions otherwise (the smoothing
parameter was chosen subjectively and the same value was used in all cases).
The $\phi _{k}$s and $\psi _{l}$s were modeled as cubic B-splines with nine
equally spaced knots, as before. Overall, we considered five combinations $%
(p,q,r)$: $(1,1,1)$, $(1,1,2)$, $(2,2,2)$, $(2,2,3)$, and $(3,3,3)$; the
model closest to the truth is $(2,2,2)$.

\begin{table}[tbp] \centering%
\begin{tabular}{cccccc}
& \multicolumn{2}{c}{$\mathrm{MIAE}(\hat{\beta})$} &  & \multicolumn{2}{c}{$%
\mathrm{MIAE}(\hat{w})\times 10$} \\ 
$(p,q,r)$ & H & MS &  & H & MS \\ 
$(1,1,1)$ & .637 & .226 &  & .171 & .108 \\ 
$(1,1,2)$ & .220 & .215 &  & .068 & .081 \\ 
$(2,2,2)$ & .150 & .138 &  & .066 & .079 \\ 
$(2,2,3)$ & .215 & .171 &  & .081 & .084 \\ 
$(3,3,3)$ & .231 & .184 &  & .081 & .084%
\end{tabular}%
\caption{Simulation results for Model 2. Mean integrated absolute errors
of the slope estimators $\hat{\beta}$ and the warping functions are given, for dynamic regression estimators using 
Hermite splines (H) or monotone smooth transformations (MS) as warping functions.}
\label{tab:Sim_Mod2}%
\end{table}%

In addition to the mean integrated absolute errors of the $\hat{\beta}$s we
wanted to assess the warping quality, so we also computed $\mathrm{MIAE}(%
\hat{w})=\mathrm{E}\{n^{-1}\sum_{i=1}^{n}\int |\hat{w}_{i}(t)-w_{i}(t)|dt\}$%
. They are shown in Table \ref{tab:Sim_Mod2}. We see that the optimal model
is $(2,2,2)$ as expected, and that monotone smooth transformations generally
produce smaller estimation errors for $\beta $ than Hermite splines,
although the latter produce smaller warping errors, probably because the
true warping functions were also splines. Monotone smooth transformations
seem to be more robust to misspecification of the warping knots, although
this comes at the price of having to select a smoothing parameter.

\section{\label{sec:Example}Application: Lip Movement Data}

In this section we apply the new estimation method to the data of Malfait
and Ramsay (2003). As explained in the Introduction, the goal is to predict
lip acceleration (Figure \ref{fig:Sample}(b)) using lip neural activity
(Figure \ref{fig:Sample}(a)). This data is hard to analyze for a number of
reasons: the curves have sharp peaks and valleys, the first EMG spike occurs
very close to the origin, there is substantial phase variability, and there
are only 29 sample curves left after removing 3 obvious outliers. We
computed dynamic and ordinary retrospective regression estimators with
different numbers of components $(p,q)$ and chose the best model by
five-fold cross-validation (see Table \ref{tab:Example_CV5}). The principal
components were modeled as cubic B-splines with knots at $\{.05,.10,\ldots
,.65\}$. As warping functions we used Hermite splines with knots $\mathbf{%
\tau }_{0}=(.08,.2,.4,.5)$, which approximately correspond to the average
location of the EMG peaks.

\begin{table}[tbp] \centering%
\begin{tabular}{ccc}
$(p,q)$ & D & L \\ 
$(1,1)$ & .244 & .293 \\ 
$(2,2)$ & .223 & .275 \\ 
$(3,3)$ & .222 & .257 \\ 
$(4,4)$ & .236 & .238 \\ 
$(5,5)$ & .220 & .231 \\ 
$(6,6)$ & --- & .232 \\ 
$(7,7)$ & --- & .232%
\end{tabular}%
\caption{Lip Movement Example. Cross-validated mean prediction errors for several models of 
dynamic functional regression (D) and ordinary linear regression (L).}\label%
{tab:Example_CV5}%
\end{table}%

According to Table \ref{tab:Example_CV5}, the best dynamic regression
estimator is given by a three-component model and the best ordinary linear
regression estimator by a five-component model. Figure \ref{fig:Sample}(c,d)
shows the warped sample curves. We see that dynamic regression does a good
job at synchronizing the curves. The features of both explanatory and
response curves emerge very clearly. In particular, the peaks of the EMG
curves around $t=.4$ and $t=.5$, which were barely discernible in Figure \ref%
{fig:Sample}(a), are plain to see in Figure \ref{fig:Sample}(c). These peaks
correspond to the agonistic and antagonistic actions of the lower-lip muscle
at the beginning and the end of the second `b' in `Bob'.

Figure \ref{fig:Fitted_curves} shows the lip acceleration curves $\{y_{i}\}$
together with the fitted curves $\{\hat{y}_{i}\}$. We see that ordinary
least squares produces a substantially worse fit; the mean prediction error
of dynamic regression is .0898 while the mean prediction error of ordinary
regression is .1648, almost twice as large. Even though ordinary regression
uses two more principal components than dynamic regression to estimate $%
\beta $, it is clear that these extra components cannot make up for the lack
of a time-warping mechanism, and adding more components actually makes
prediction worse, as Table \ref{tab:Example_CV5} shows. So this is a
situation where the data clearly calls for a model that includes a
time-warping mechanism, and dynamic regression then represents a substantial
improvement over ordinary linear regression.

\FRAME{ftbpFU}{7.0785in}{1.6388in}{0pt}{\Qcb{Lip Movement Example. (a)
Response curves; (b) fitted curves obtained by dynamic regression; (c)
fitted curves obtained by ordinary linear regression.}}{\Qlb{%
fig:Fitted_curves}}{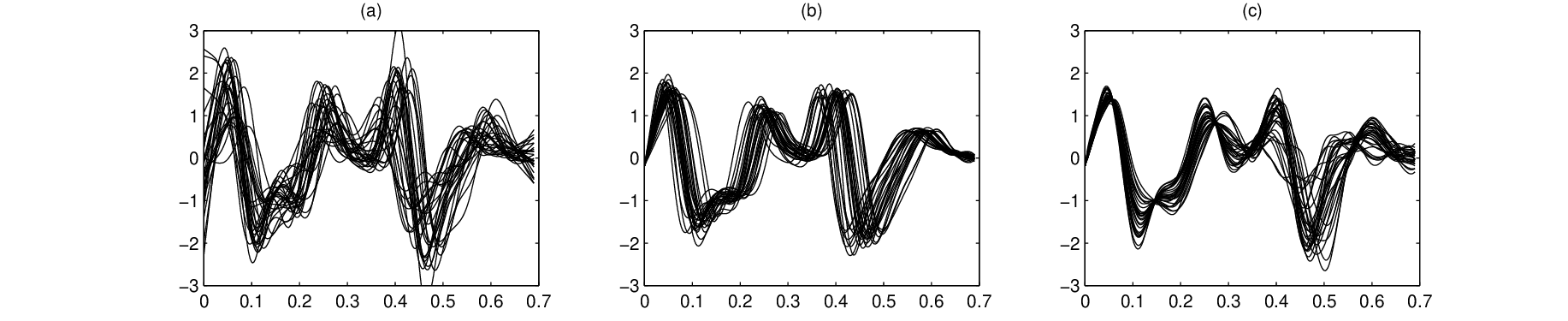}{\special{language "Scientific Word";type
"GRAPHIC";maintain-aspect-ratio TRUE;display "USEDEF";valid_file "F";width
7.0785in;height 1.6388in;depth 0pt;original-width 12.2233in;original-height
2.4768in;cropleft "0.1135";croptop "1";cropright "1";cropbottom "0";filename
'yhats.eps';file-properties "XNPEU";}}

Interpreting $\hat{\beta}(s,t)$ is harder but also interesting. To determine
which features of the $\hat{\beta}$s are actually statistically significant,
we estimated the variance of $\hat{\beta}(s,t)$, $\hat{v}(s,t)$, by
bootstrap (using residual resampling). Contour plots of the filtered
estimators $\hat{\beta}(s,t)\mathbb{I}\{|\hat{\beta}(s,t)|\geq 2\sqrt{\hat{v}%
(s,t)}\}$ are shown in Figure \ref{fig:Betas}. The dynamic regression
estimator shows significant features outside the diagonal, implying that lip
acceleration can be predicted not only by neural activity immediately
preceding the event, but also by neural activity further in the past. For
example, consider predicting the sharp deceleration of the $\tilde{y}_{i}$s
at $t=.45$, which is given by $\mu _{\tilde{y}}(.45)+\int_{0}^{.45}\beta
(s,.45)\{\tilde{x}_{i}(s)-\mu _{\tilde{x}}(s)\}ds$. In Figure \ref{fig:Betas}%
(a) we see that $\beta (s,.45)$ not only has a peak near the diagonal, which
is unsurprising because it corresponds to the immediately preceding neural
activity, but also at $s=.1$ (where the valley between the first two peaks
of the $\tilde{x}_{i}$s occur), and troughs before and after that peak
(where the first two peaks of the $\tilde{x}_{i}$s occur). This implies that
if the first two spikes of $\tilde{x}_{i}$ (related to the \emph{first} `b')
are sharper than the mean, the integral $\int_{0}^{.45}\beta (s,.45)\{\tilde{%
x}_{i}(s)-\mu _{\tilde{x}}(s)\}ds$ will tend to be negative and then $\tilde{%
y}_{i}(.45)$, the deceleration of the lips at the \emph{second} `b', will
tend to be stronger than the mean. Off-diagonal features of the ordinary
least squares estimator can also be seen in Figure \ref{fig:Betas}(b), and
were also observed by Malfait and Ramsay (2003) using a different approach
to ordinary least squares (based on a triangular-basis expansion for $\beta $
rather than on a tensor-product principal-component expansion), but they are
harder to interpret because they are applied to non-synchronized curves.

\FRAME{ftbpFU}{6.5717in}{4.1035in}{0pt}{\Qcb{Lip Movement Example. Contour
plots of estimated slope functions $\hat{\protect\beta}(s,t)$ [(a) dynamic
regression estimator, (b) ordinary least squares estimator].}}{\Qlb{fig:Betas%
}}{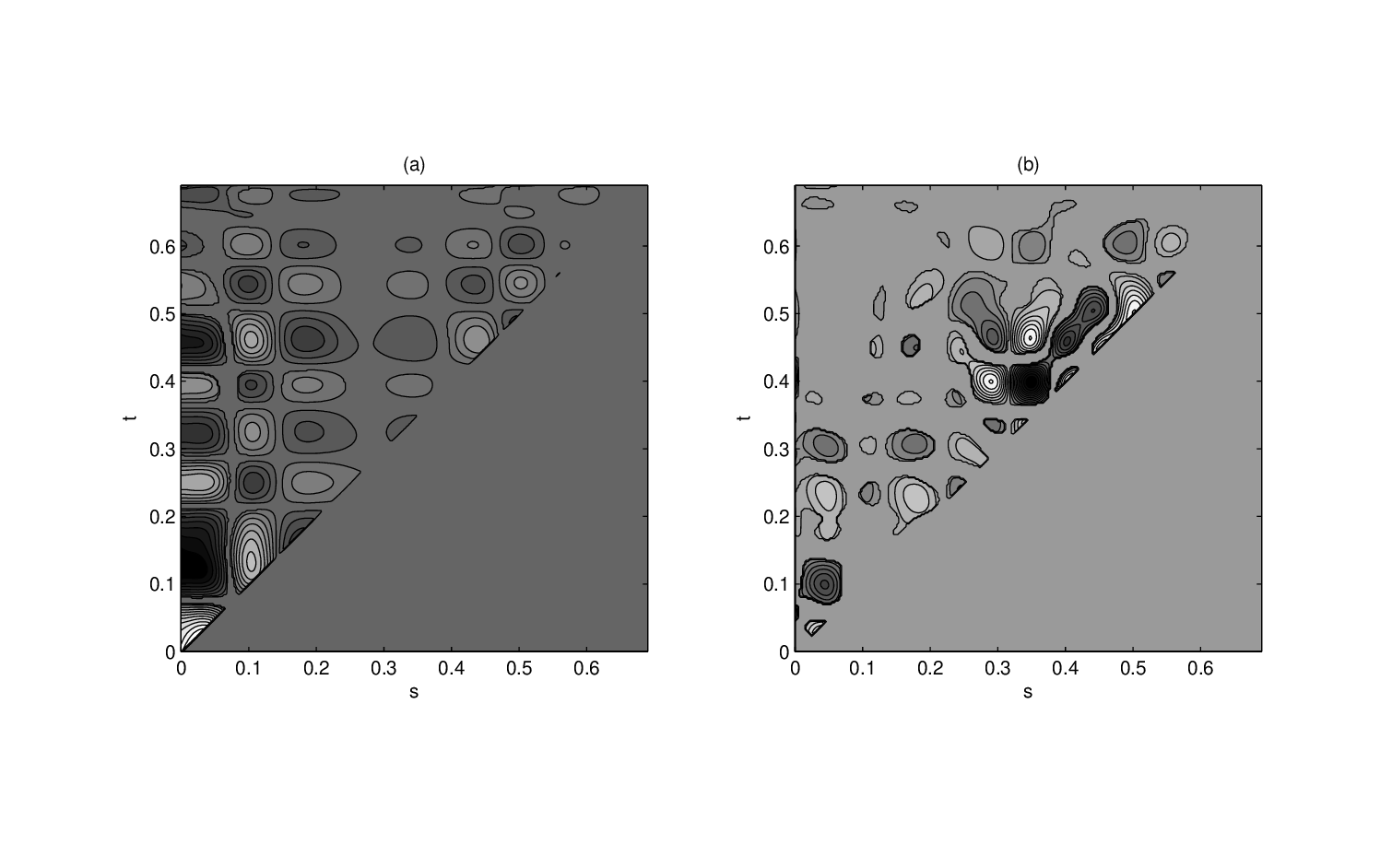}{\special{language "Scientific Word";type
"GRAPHIC";maintain-aspect-ratio TRUE;display "USEDEF";valid_file "F";width
6.5717in;height 4.1035in;depth 0pt;original-width 9.1506in;original-height
6.4489in;cropleft "0";croptop "1";cropright "1";cropbottom "0";filename
'Betas.eps';file-properties "XNPEU";}}

More information about the dynamics of the process can be extracted from the
warping functions themselves. The use of interpolating Hermite splines
facilitates this, because the estimated parameters $\mathbf{\hat{\tau}}_{i}$
roughly correspond to the locations of the landmarks $\mathbf{\tau }_{0}$ on
the respective sample curve. For our choice of $\mathbf{\tau }_{0}$, the $%
\mathbf{\hat{\tau}}_{i}$s will roughly correspond to the location of the
four characteristic peaks of the EMG curves. Thus $d_{i1}=\hat{\tau}_{i2}-%
\hat{\tau}_{i1}$ indicates the duration of the first `b', $d_{i2}=\hat{\tau}%
_{i3}-\hat{\tau}_{i2}$ the duration of the `o', and $d_{i3}=\hat{\tau}_{i4}-%
\hat{\tau}_{i3}$ the duration of the second `b'. The pairwise correlations
of the $d$s are $\rho _{12}=-.46$, $\rho _{13}=.66$ and $\rho _{23}=-.24$,
indicating that there is a significant negative correlation between the
duration of the first `b' and the `o', and a significant positive
correlation between the durations of the two `b's. More accurate information
about the phonemes' duration could be obtained by estimating the exact peak
locations curve by curve, but that would be unfeasible for larger datasets.
An advantage of Hermite-spline warping is that the $\mathbf{\tau }_{i}$s are
estimated automatically as a by-product of the procedure.

\section*{Acknowledgement}

This research was partially supported by NSF grant DMS 10-06281.

\section*{References}

\begin{description}
\item Brumback, L.C.~and Lindstrom, M.J. (2004). Self modeling with
flexible, random time transformations. \emph{Biometrics }\textbf{60 }%
461--470.

\item Cai, T. and Hall, P. (2006). Prediction in functional linear
regression. \emph{The Annals of Statistics} \textbf{34} 2159--2179.

\item Crambes, C., Kneip, A., and Sarda, P. (2009). Smoothing splines
estimators for functional linear regression. \emph{The Annals of Statistics }%
\textbf{37} 35--72.

\item Fritsch, F.N.~and Carlson, R.E. (1980). Monotone piecewise cubic
interpolation. \emph{SIAM Journal of Numerical Analysis }\textbf{17}
238--246.

\item Gervini, D. and Gasser, T. (2004). Self-modeling warping functions. 
\emph{Journal of the Royal Statistical Society (Series B)} \textbf{66}
959--971.

\item Gervini, D. and Gasser, T. (2005). Nonparametric maximum likelihood
estimation of the structural mean of a sample of curves. \emph{Biometrika} 
\textbf{92} 801--820.

\item Gohberg, I., Goldberg, S., and Kaashoek, M. A. (2003). \emph{Basic
Classes of Linear Operators}. Basel: Birkh\"{a}user Verlag.

\item Hall, P.~and Horowitz, J. L. (2007). Methodology and convergence rates
for functional linear regression. \emph{The Annals of Statistics} \textbf{35}
70--91.

\item Hastie, T., Tibshirani, R. and Friedman, J. (2009). \emph{The Elements
of Statistical Learning. Data Mining, Inference, and Prediction. Second
Edition.} Springer, New York.

\item Hurvich, C.M., Simonoff, J.S., and Tsai, C.-L. (1998). Smoothing
parameter selection in nonparametric regression using an improved Akaike
information criterion. \emph{Journal of the Royal Statistical Society
(Series B)} \textbf{60} 271--293.

\item James, G., Wang, J. and Zhu, J. (2009). Functional linear regression
that's interpretable. \emph{The Annals of Statistics} \textbf{37} 2083--2108.

\item James, G.M. (2007). Curve alignment by moments. \emph{The Annals of
Applied Statistics} \textbf{1} 480--501.

\item Jupp, D. L. B. (1978). Approximation to data by splines with free
knots. \emph{SIAM J. Numer. Anal.} \textbf{15} 328--343.

\item Kneip, A., Li, X., MacGibbon, B. and Ramsay, J.O. (2000). Curve
registration by local regression. \emph{Canadian Journal of Statistics} 
\textbf{28} 19--30.

\item Kneip, A. and Ramsay, J.O. (2008). Combining registration and fitting
for functional models. \emph{Journal of the American Statistical Association}
\textbf{103} 1155--1165.

\item Liang, H., Wu, H., and Carroll, R. J. (2003). The relationship between
virologic and immunologic responses in AIDS clinical research using
mixed-effects varying-coefficient models with measurement error. \emph{%
Biostatistics} \textbf{4} 297--312.

\item Liu, X. and M\"{u}ller, H.-G. (2004). Functional convex averaging and
synchronization for time-warped random curves. \emph{Journal of the American
Statistical Association} \textbf{99} 687--699.

\item Malfait, N. and Ramsay, J.O. (2003). The historical functional linear
model. \emph{Canadian Journal of Statistics} \textbf{31} 115--128.

\item M\"{u}ller, H.-G., Chiou, J.-M., and Leng, X. (2008). Inferring gene
expression dynamics via functional regression analysis. \emph{BMC
Bioinformatics} \textbf{9} 60.

\item Nocedal, J.~and Wright, S.J. (2006). \emph{Numerical Optimization.
Second Edition}. Springer, New York.

\item Ramsay, J.O. and Li, X. (1998). Curve registration. \emph{Journal of
the Royal Statistical Society (Series B)} \textbf{60} 351--363.

\item Ramsay, J.O. and Silverman, B. (2005). \emph{Functional Data Analysis.
Second Edition.} Springer, New York.

\item Silverman, B. (1996). Smoothed functional principal components
analysis by choice of norm. \emph{The Annals of Statistics} \textbf{24}
1--24.

\item Tang, R., and M\"{u}ller, H.-G. (2008). Pairwise curve synchronization
for functional data. \emph{Biometrika} \textbf{95} 875--889.

\item Tang, R., and M\"{u}ller, H.-G. (2009). Time-syncronized clustering of
gene expression trajectories. \emph{Biostatistics} \textbf{10} 32--45.

\item Yao, F., M\"{u}ller, H.-G. and Wang, J.-L. (2005). Functional linear
regression analysis for longitudinal data. \emph{The Annals of Statistics }%
\textbf{33 }2873--2903.

\item Wahba, G. (1990). \emph{Spline Models for Observational Data.}
CBMS-NSF regional conference series in applied mathematics, SIAM,
Philadelphia.

\item Wang, K. and Gasser, T. (1999). Synchronizing sample curves
nonparametrically. \emph{The Annals of Statistics} \textbf{27 }439--460.

\item Wu, H. and Liang, H. (2004). Backfitting random varying-coefficient
models with time-dependent smoothing covariates. \emph{Scandinavian Journal
of Statistics} \textbf{31} 3--19.

\item Wu, S.~and M\"{u}ller, H.-G. (2010). Response-additive regression for
longitudinal data. \emph{Unpublished manuscript.} Available at
http://anson.ucdavis.edu/\symbol{126}mueller/rare8.pdf.
\end{description}

\end{document}